\documentclass[%
 aip,
 amsmath,amssymb,
 reprint,%
]{revtex4-1}

\usepackage{graphicx}
\usepackage{dcolumn}
\usepackage{bm}

\usepackage[utf8]{inputenc}
\usepackage[T1]{fontenc}
\usepackage{mathptmx}
\usepackage{etoolbox}

\makeatletter
\def\@email#1#2{%
 \endgroup
 \patchcmd{\titleblock@produce}
  {\frontmatter@RRAPformat}
  {\frontmatter@RRAPformat{\produce@RRAP{*#1\href{mailto:#2}{#2}}}\frontmatter@RRAPformat}
  {}{}
}%
\makeatother
\begin{document}

\preprint{AIP/123-QED}

\title{Coherent transfer via parametric control of normal-mode splitting in a superconducting multimode resonator}
\author{Kai-I Chu*}
\email[Author to whom correspondence should be addressed: ]{kaiichu0903@gmail.com}
\affiliation{Department of Physics, National Central University, Taoyuan City 32001, Taiwan}

\author{Xiao-Cheng~Lu}
\affiliation{Department of Physics, National Tsing Hua University, Hsinchu 30013, Taiwan}
\author{Hsin~Chang }
\affiliation{Department of Physics, National Central University, Taoyuan City 32001, Taiwan}
\author{Wei-Cheng~Hung}
\affiliation{Department of Physics, National Central University, Taoyuan City 32001, Taiwan}
\author{Jing-Yang~Chang}
\affiliation{Department of Physics, National Central University, Taoyuan City 32001, Taiwan}
\author{Jeng-Chung~Chen}
\affiliation{Department of Physics, National Tsing Hua University, Hsinchu 30013, Taiwan}
\author{Chii-Dong Chen}
\affiliation{Institute of Physics, Academia Sinica, Taipei 11529, Taiwan}
\author{Yung-Fu~Chen}
\affiliation{Department of Physics, National Central University, Taoyuan City 32001, Taiwan}
\affiliation{Quantum Technology Center, National Central University, Taoyuan City 32001, Taiwan}
\affiliation{Center for High Energy and High Field Physics, National Central University, Taoyuan City 32001, Taiwan}
\affiliation{Taiwan Semiconductor Research Institute, Hsinchu 30013, Taiwan}

\date{\today}

\begin{abstract}
Microwave storage and retrieval are essential capabilities for superconducting quantum circuits. Here, we demonstrate an on-chip multimode resonator in which strong parametric modulation induces a large and tunable normal-mode splitting that enables microwave storage. When the spectral bandwidth of a short microwave pulse covers the two dressed-state absorption peaks, part of the pulse is absorbed and undergoes coherent energy exchange between the modes, producing a clear time-domain beating signal. By switching off the modulation before the beating arrives, we realize on-demand storage and retrieval, demonstrating an alternative approach to microwave photonic quantum memory. This parametric-normal-mode-splitting protocol offers a practical route toward a controllable quantum-memory mechanism in superconducting circuits.
\end{abstract}

\maketitle

Parametric modulation is a promising method for enabling interactions between two detuned harmonic oscillators \cite{Dobrindt2008,Teufel2011, Okamoto2013,ZakkaBajjani2011,Sirois2015,Bothner2021,Liu2023}. By periodically modulating system parameters, one can induce frequency-mixing effects that generate an effective coupling between them. For example, in superconducting circuits, a superconducting quantum interference device (SQUID) is often modulated by an AC flux drive \cite{ZakkaBajjani2011,Sirois2015,Bothner2021}. When the modulation frequency nearly matches the detuning between two modes, the interaction is activated through parametric control. Intuitively, the modulation hybridizes the two modes and creates a normal-mode splitting between them.
For quantum information processing, previous works have demonstrated frequency-conversion-based storage between detuned resonators \cite{Sirois2015,Liu2023}, electromagnetically induced transparency (EIT)–based storage in qubit–resonator systems \cite{Chu2025}, and the implementation of two-qubit gates \cite{Strand2013,McKay2016,Didier2018,Caldwell2018,Li2018} to generate entangled states. However, strong parametric modulation that induces a large normal-mode splitting to coherently transfer microwave signals between different modes has not been systematically studied. In particular, this strategy is analogous to the Autler–Townes-splitting (ATS) quantum memory protocol \cite{Saglamyurek2018,Saglamyurek2019} in atomic systems. Such an approach offers a promising way to store short-duration pulses by using the dressed-state absorption peaks, in contrast to the EIT protocol \cite{Liu2001,Fleischhauer2005,Hsiao2018}. To the best of our knowledge, storage based on parametrically controlled normal-mode splitting has not been demonstrated previously in superconducting circuits. This work demonstrates such coherent transfer via quantum beating between two normal-mode-splitting absorption peaks in a multimode resonator system. By dynamically turning the parametric control off and on, the beating signal can be retrieved on demand, similar to the ATS-based storage-and-retrieval dynamics. Our approach provides a new method for on-chip pulse storage in superconducting circuits.

\begin{figure}[t!]
    \centering
    \includegraphics[width=0.49\textwidth]{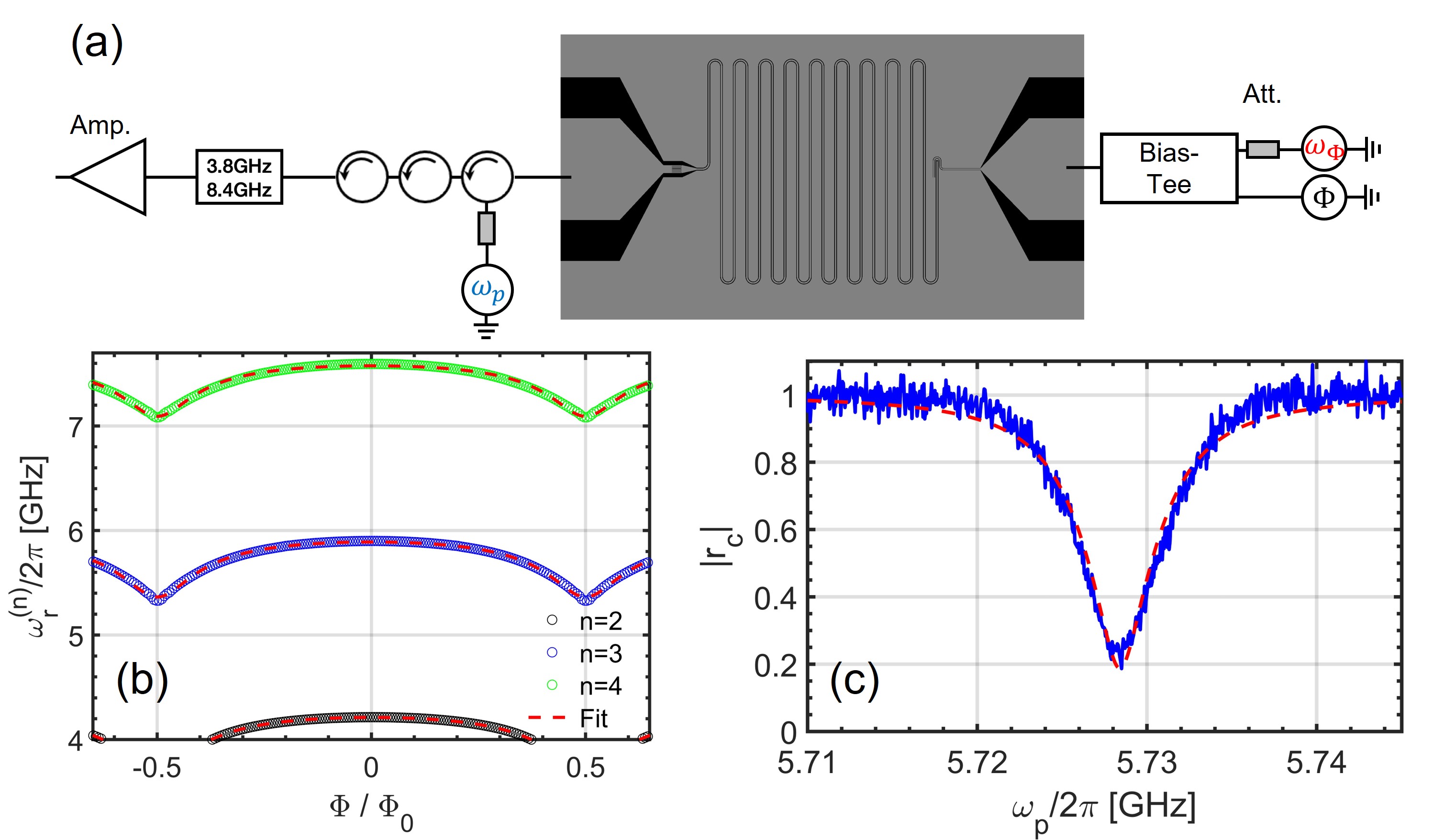}
    \caption{(a) Optical micrograph of the resonator together with a diagram of the measurement setup. The symbols $\omega_\Phi$, $\Phi$, and $\omega_p$ denote the parametric modulation tone, the DC flux bias, and the probe tone, respectively. Att. and Amp. denote the attenuator and the amplifier. (b) Resonance frequency of each mode $\omega_r^{(n)}$ as a function of the DC flux $\Phi$. The red dashed lines represent the theoretical fits. (c) Reflection coefficient $|r_c|$ (blue) as a function of the probe frequency $\omega_p$ when the resonator is biased at $\Phi = 0.33\Phi_0$. The red dashed line denotes the theoretical fit.}
    \label{fig:setup}
\end{figure}

Our circuit and setup, illustrated in Fig. \ref{fig:setup}(a), is a tunable $\lambda/4$ resonator with one end capacitively coupled to the output port and the other end terminated to ground through a SQUID. The fundamental mode of the resonator is designed to be approximately $\omega_r^{(0)}/2\pi = 900~\mathrm{MHz}$ when the SQUID is replaced by a short to ground. Ideally, the resonance frequencies of the higher‐order modes follow $\omega_r^{(n)} = (2n+1)\omega_r^{(0)}$. However, because the terminating SQUID provides a different effective inductance to each mode, their actual resonance frequencies deviate slightly from this ideal relation \cite{ZakkaBajjani2011}. This mode separation allows us to selectively activate the desired interaction through parametric modulation. The chip is cooled in a dilution refrigerator to a base temperature of about 20 mK. A DC flux bias is combined with a parametric modulation tone at frequency $\omega_\Phi$ using a bias tee to control the SQUID inductance and enable coupling between the modes. A probe tone at frequency $\omega_p$ is sent into the resonator, reflected, and routed through a circulator, filter, and amplifier before being demodulated for measurement. The measurable frequencies of our setup span from 4 to 8 GHz.

By fitting the reflection coefficient $|r_c|$ of the resonator in Fig. \ref{fig:setup}(c), the resonance frequencies and the loss rates of each mode can be extracted. The probe power for the spectroscopy experiments is fixed at $P_p=-117$ dBm. The resonance frequency of the $n$-th mode can be expressed as $\omega^{(n)}_r=2\pi/\sqrt{C_n(L_n+L_s(\Phi))}$, where $L_n$ and $C_n$ are the effective inductance and capacitance of the mode, and $L_s(\Phi)=\Phi_0/(4\pi I_c\sqrt{\cos^2(\pi\Phi/\Phi_0)+d^2\sin^2(\pi\Phi/\Phi_0)})$ is the flux-dependent SQUID inductance contributed to the resonator. Here, $I_c$ is the critical current, $d$ is the junction asymmetry, $\Phi$ is the applied flux, and $\Phi_0$ is the flux quantum. The resulting $n$-th mode frequencies as a function of $\Phi$ are plotted in Fig.~\ref{fig:setup}(b). The resonator is biased at $\Phi = 0.33\Phi_0$ so that the resonance frequencies vary approximately linearly with flux, which is favorable for parametric modulation. At this bias, the extracted resonance frequencies are $\omega_r^{(2)}/2\pi = 4.0614~\mathrm{GHz}$, $\omega_r^{(3)}/2\pi = 5.7284~\mathrm{GHz}$, and $\omega_r^{(4)}/2\pi = 7.4203~\mathrm{GHz}$. The corresponding total loss rates are $\kappa_{\mathrm{tot}}^{(2)}/2\pi = 4.6461~\mathrm{MHz}$, $\kappa_{\mathrm{tot}}^{(3)}/2\pi = 6.8857~\mathrm{MHz}$, and $\kappa_{\mathrm{tot}}^{(4)}/2\pi = 8.3224~\mathrm{MHz}$. The difference between the adjacent mode spacings, $(\omega_r^{(3)} - \omega_r^{(2)}) - (\omega_r^{(4)} - \omega_r^{(3)})$, is slightly larger than the bandwidth of each mode, ensuring that the parametric drive selectively couples \cite{ZakkaBajjani2011} only modes 2 and 3 in the subsequent experiment.

\begin{figure}[t!]
    \centering
    \includegraphics[width=0.49\textwidth]{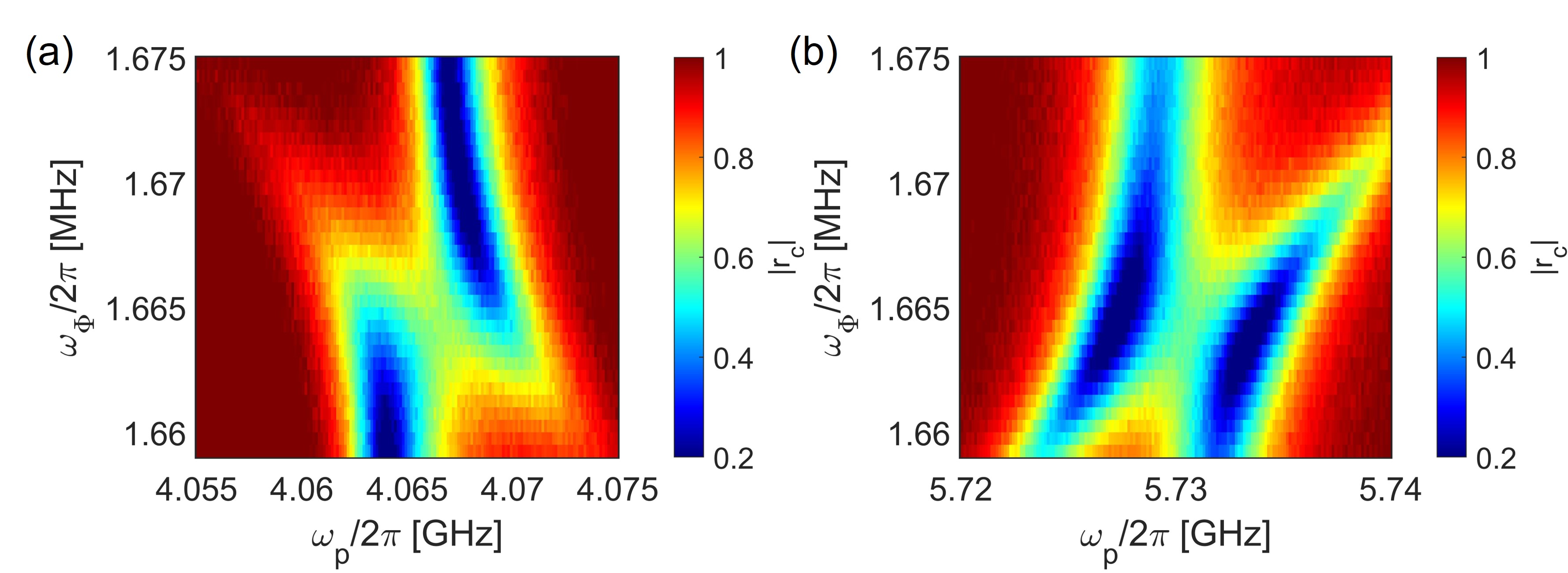}
    \caption{Normal-mode–splitting spectrum under parametric modulation with a fixed flux-modulation amplitude. $|r_c|$ is plotted as a function of the modulation frequency $\omega_\Phi$ and the probe frequency $\omega_p$, where $\omega_p$ is tuned near resonance with (a) mode 2 and (b) mode 3.}
    \label{fig:splitting_dir}
\end{figure}

By applying a flux modulation at frequency $\omega_\Phi \approx \omega_r^{(3)} - \omega_r^{(2)}$ with amplitude $\delta\Phi$, an effective coupling strength $g_\Phi$ is induced between modes 2 and 3, producing a clear normal-mode splitting in the spectrum, as shown in Fig. \ref{fig:splitting_dir}. The direction in which the two resonance branches split reveals whether a mode hybridizes with a higher- or lower-frequency partner. The opposite splitting orientations observed in Fig.~\ref{fig:splitting_dir}(a) and (b) confirm that mode 2 is coupled to mode 3.

\begin{figure}[t!]
    \centering
    \includegraphics[width=0.49\textwidth]{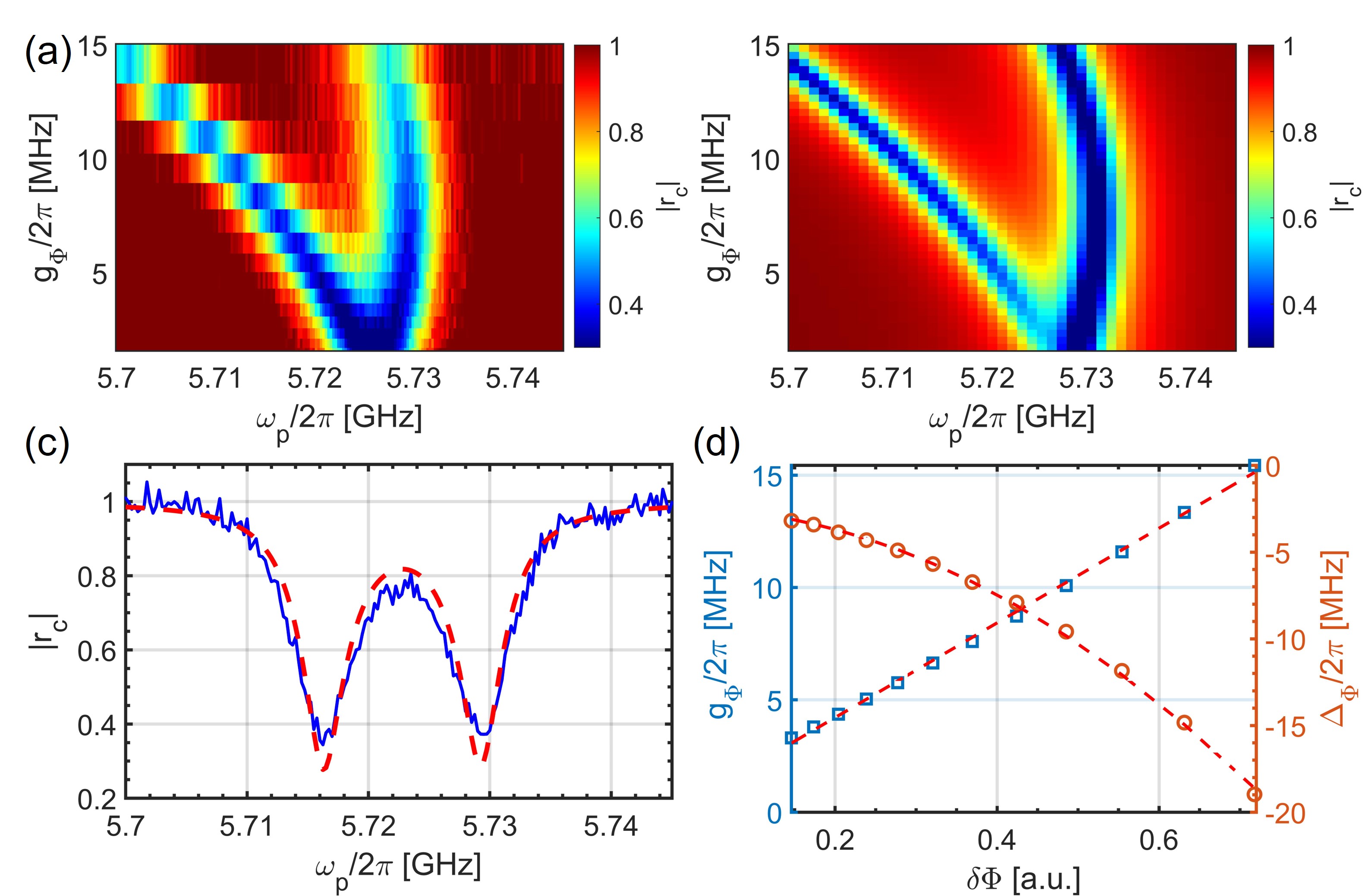}
    \caption{(a) False-color spectroscopy of $|r_c|$ and (b) the corresponding simulation, plotted as functions of $g_\Phi$ and the probe frequency $\omega_p$. The modulation frequency $\omega_\Phi/2\pi=1.664$ is chosen near the detuning between modes 2 and 3. (c) Line cut of the experimental data at $g_\Phi/2\pi=6.6$ MHz, which is denoted by the blue line. The red dashed curve indicates the theoretical fit. (d) Extracted $g_\Phi$ and detuning $\Delta_\Phi$ as functions of the flux-modulation amplitude $\delta\Phi$. The empty circles represent the fitting results from (a). The red dashed lines show the linear fit for $g_\Phi$ and the quadratic fit for $\Delta_\Phi$.}
    \label{fig:splitting}
\end{figure}

The Hamiltonian of the two coupled resonator modes under parametric flux modulation and probe field, within the rotating-wave approximation, can be expressed as 
\begin{equation}
\begin{aligned}
H=&(\tilde{\omega}^{(3)}_r-\omega_p)a^\dagger a + (\tilde{\omega}^{(2)}_r-\omega_p+\omega_\Phi)b^\dagger b
\\&+g_\Phi(a^\dagger b + b^\dagger a) +\sqrt{\kappa^{(3)}_{ext}}\alpha_{in}(a^\dagger +a),
\end{aligned}
\end{equation}
where $a(a^\dagger)$ and $b(b^\dagger)$ are the annihilation (creation) operators of modes 3 and 2, respectively. The coherent probe field is denoted by $\alpha_{in}$, and the external loss rate of mode 3 is $\kappa^{(3)}_{ext}/2\pi=4.0874$ MHz. Although the resonator is biased close to the linear-response region of $\omega_r^{(n)}(\Phi)$, the small nonlinearity visible in Fig.~\ref{fig:setup}(b) leads to a motional-averaging effect under flux modulation. This results in slight shifts of the mode resonance frequencies \cite{Beaudoin2012}, which we denote as $\tilde{\omega}_r^{(3)}$ and $\tilde{\omega}_r^{(2)}$. The modulation-induced coupling strength is linearly proportional to the flux modulation amplitude $g_\Phi\propto\delta\Phi$ \cite{Sirois2015}. The corresponding Langevin equations are
\begin{equation}
\begin{aligned}
\dot{a}=[-i(\tilde{\omega}^{(3)}_r-\omega_p)-\frac{\kappa^{(3)}_{tot}}{2}]a-ig_\Phi b-i\sqrt{\kappa^{(3)}_{ext}}\alpha_{in},
\end{aligned}
\end{equation}
\begin{equation}
\begin{aligned}
\dot{b}=[-i(\tilde{\omega}^{(2)}_r-\omega_p+\omega_\Phi)-\frac{\kappa^{(2)}_{tot}}{2}]b-ig_\Phi a.
\end{aligned}
\end{equation}
The output probe field $\alpha_{out}$ follows the standard input–output relation
\begin{equation}
\begin{aligned}
\alpha_{out}=\alpha_{in}+i\sqrt{\kappa^{(3)}_{ext}}a.
\end{aligned}
\end{equation}

We further study the steady state spectrum by fixing the modulation frequency near the detuning between modes 2 and 3 at $\omega_\Phi/2\pi=1.664$ GHz. The experimental data and numerical simulations are shown in Fig. \ref{fig:splitting} (a) and (b). The simulations are obtained by solving the steady-state Langevin equations together with the input–output relation to evaluate the reflection coefficient $r_c=\alpha_{\mathrm{out}}/\alpha_{\mathrm{in}}$, and they reproduce the experimental spectra well. As the modulation amplitude increases, thereby enhancing $g_\Phi$, the magnitude of the splitting increases accordingly. A slight shift in the resonance frequencies is also observed, originating from the motional-averaging effect discussed above. The steady-state optical response of a driven $\Lambda$-type system embedded at the end of the transmission line can be used to fit the data in Fig.~\ref{fig:splitting}(a). Its analytical form \cite{Chu2025,Gu2017,Chu2023,Chu2025_2} is 
 \begin{equation}
    \label{eq:Driven Lambda type}
    r^{a}_{c} = 1 + 2i\frac{\frac{\kappa^{(3)}_{ext}}{2}({\delta-i\frac{\kappa^{(2)}_{tot}}{2})}}{(\delta-i\frac{\kappa^{(2)}_{tot}}{2})(\delta+\Delta_2-i\frac{\kappa^{(3)}_{tot}}{2})-g_\Phi^2},
\end{equation}
where $\delta=\Delta_1-\Delta_2$, $\Delta_1=\tilde{\omega}^{(3)}_{r}-\omega_p$, and $\Delta_2=\tilde{\omega}^{(3)}_{r}-\tilde{\omega}^{(2)}_{r}-\omega_{\Phi}$.
A line cut of the data in Fig.~\ref{fig:splitting}(a) is well fitted by Eq.~\ref{eq:Driven Lambda type}, as shown in Fig.~\ref{fig:splitting}(c). The extracted values of $g_\Phi$ and the detuning
$\Delta_\Phi = \tilde{\omega}_r^{(3)} - \omega_r^{(3)}$
are plotted as functions of $\delta\Phi$ in Fig.~\ref{fig:splitting}(d). As expected, $g_\Phi$ increases linearly with $\delta\Phi$. In addition, the scaling of $\Delta_\Phi$ follows a quadratic dependence on $\delta\Phi$.

\begin{figure}[t!]
    \centering
    \includegraphics[width=0.49\textwidth]{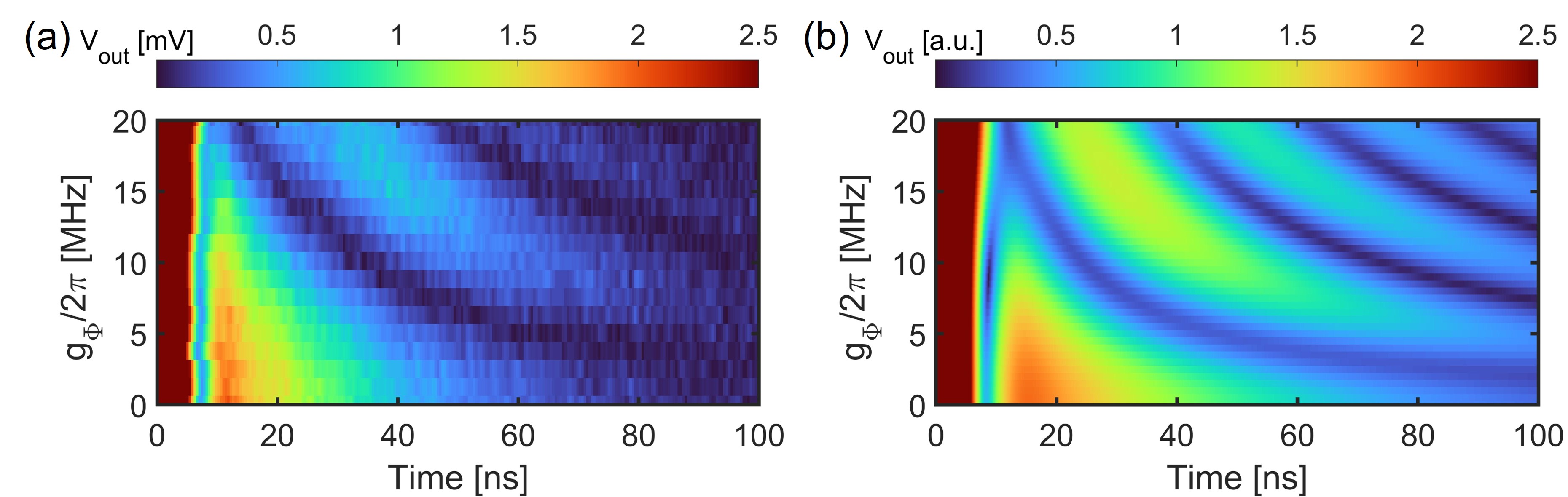}
    \caption{A short Gaussian pulse is injected into the resonator with a fixed modulation frequency $\omega_\Phi/2\pi=1.664$ GHz while varying the coupling strength $g_\Phi$. The measured output voltage $V_{\mathrm{out}}$ is plotted as a function of $g_\Phi$ and time. (a) Experimental results. (b) Simulation results.}
    \label{fig:beating}
\end{figure}

The normal-mode splitting induced by the strong parametric coupling between modes 2 and 3 implies that a probe pulse can be partially absorbed by the two dressed-state resonances, provided that its spectral bandwidth covers the splitting. The absorbed component is then coherently exchanged between the two modes, giving rise to a time-domain beating signal. To observe this effect, a short Gaussian probe pulse with carrier frequency $\omega_p/2\pi=5.728$ GHz and envelope $\exp(-t^2/{\tau_d}^2)$, where $\tau_d=5$ ns and $P_p=-102$ dBm, is applied to the resonator under parametric modulation. Figure~\ref{fig:beating} shows the resulting time-domain traces for different values of $g_\Phi$ with fixed $\omega_\Phi/2\pi=1.664$ GHz. When $g_\Phi$ is too small for the splitting to be resolved, the output exhibits a simple exponential decay from the resonator emission.
As $g_\Phi$ increases, the two dressed modes absorb a portion of the pulse, and a clear beating pattern emerges, indicating coherent energy exchange between modes 2 and 3. The simulations show the same behavior and agree well with the experimental data.

\begin{figure}[tp]
    \centering
    \includegraphics[width=0.4\textwidth]{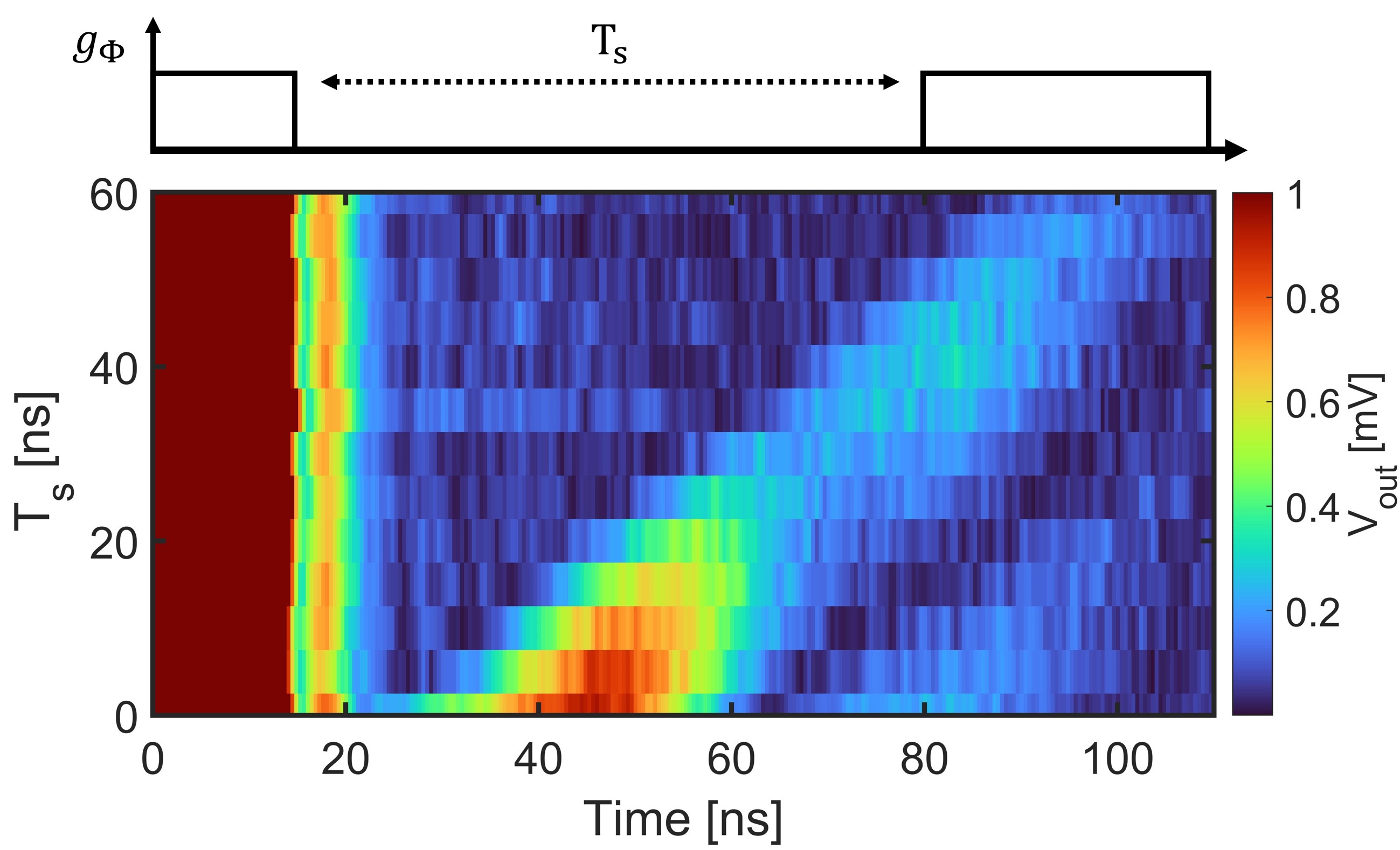}
    \caption{On-demand storage and retrieval of the beating signal. The measured output voltage $V_{\mathrm{out}}$ is plotted as a function of the storage time $T_s$ and time. The upper panel shows the control sequence of the parametric modulation.}
    \label{fig:beating2}
\end{figure}

To operate the system as a memory, the parametric modulation is turned off before the beating signal is retrieved. $g_\Phi/2\pi=17.6$ MHz is used in the experiment. Figure~\ref{fig:beating2} shows the on-demand storage and retrieval of the beating signal by switching the modulation off and on with a storage time $T_s$. The beating amplitude decays with increasing $T_s$ due to the limited coherence time of mode 2. This storage protocol is analogous to the ATS quantum memory in atomic systems.

To improve the storage efficiency and fidelity, the modulation can, in principle, be switched off using a pulse shape matched to the Gaussian probe pulse, ensuring full spectral overlap with the dressed-state resonances \cite{Saglamyurek2018,Saglamyurek2019}. However, such dynamic control in our device induces additional frequency shifts arising from the motional-averaging effect. This limitation may be overcome by implementing a Superconducting Nonlinear Asymmetric Inductive eLement (SNAIL) \cite{Frattini2017,Sivak2019,Cao2024} to suppress the nonlinear flux dependence and eliminate the undesired dynamical shifts.

In conclusion, we demonstrate an alternative approach to photonic quantum memory in superconducting circuits by exploiting parametric-modulation–induced normal-mode splitting. We achieve microwave storage and on-demand retrieval in a multimode resonator by dynamically controlling the induced coupling between the modes. Future implementations in higher-coherence devices could further enhance the storage efficiency. Moreover, replacing the SQUID with a SNAIL element would suppress nonlinear flux dependence and enable pulse-shape–matched quantum-memory protocols.

\begin{acknowledgments}
The work was supported by the National Science and Technology Council in Taiwan through Grants No. NSTC 113-2112-M-008-006 and
NSTC 114-2112-M-008-022.
\end{acknowledgments}



\section*{AUTHOR DECLARATIONS}

\subsection*{Conflict of Interest}
The authors have no conflicts to disclose.

\subsection*{Author Contributions}
Kai-I~Chu and Xiao-Cheng~Lu contributed equally to this work.

\textbf{Kai-I Chu}: 
Conceptualization (lead); Methodology (equal); Software (lead); Formal analysis (lead); Visualization (lead); Data curation (equal); Writing – original draft (lead); Writing – review \& editing (lead); Supervision (equal).
\textbf{Xiao-Cheng Lu}: 
Methodology (equal); Investigation (lead); Validation (lead); Formal analysis (supporting); Data curation (equal); Visualization (supporting); Writing – review \& editing (equal).
\textbf{Hsin Chang}: 
Investigation (supporting); Software (supporting).
\textbf{Wei-Cheng Hung}: 
Methodology (supporting).
\textbf{Jing-Yang Chang}: 
Methodology (supporting).
\textbf{Jeng-Chung Chen}: 
Resources (supporting); Supervision (supporting).
\textbf{Chii-Dong Chen}: 
Resources (supporting); Supervision (supporting).
\textbf{Yung-Fu Chen}: 
Conceptualization (supporting); Writing – review \& editing (equal); 
Supervision (lead); Resources (lead); Funding acquisition (lead).

\subsection*{DATA AVAILABILITY}
The data that support the findings of this study are available within the article.

\nocite{*}
\bibliography{aipsamp}%

\end{document}